\documentclass[num-refs]{nbdt-article}

\usepackage{siunitx}
\usepackage{bm}
\usepackage{wrapfig}
\usepackage{amsmath}
\usepackage{graphicx}
\usepackage{subcaption}
\usepackage{caption}

\papertype{Original Article}
\paperfield{Neural data science / analysis}

\title{Synthesizing Speech from Intracranial Depth Electrodes using an Encoder-Decoder Framework}

\author[1]{Jonas Kohler}
\author[2]{Maarten Ottenhoff}
\author[2]{Sophocles Goulis}
\author[3]{Miguel Angrick}
\author[5]{Albert Colon}
\author[5]{Louis Wagner}
\author[5]{Simon Tousseyn}
\author[2,4]{Pieter L. Kubben}
\author[2]{Christian Herff}

\affil[1]{Department of Computer Science, ETH Zurich, Switzerland}
\affil[2]{Department of Neurosurgery, School of Mental Health and Neurosciences, Maastricht University, Maastricht, Netherlands}
\affil[3]{Cognitive Systems Lab, University of Bremen, Bremen, Germany}
\affil[4]{Academic Center for Epileptology, Kempenhaeghe/Maastricht University Medical Center, location Maastricht, Netherlands}
\affil[5]{Academic Center for Epileptology, Kempenhaeghe/Maastricht University Medical Center, location Kempenhaeghe, Netherlands}
\corraddress{Christian Herff, Maastricht University}
\corremail{c.herff@maastrichtuniversity.nl}

\fundinginfo{C.H. acknowledges funding by the Dutch Research Council (NWO) through the research project 'Decoding Speech In SEEG'.}
\runningauthor{Kohler et al.}

\begin{document}

\maketitle

\begin{abstract}

Speech Neuroprostheses have the potential to enable communication for people with dysarthria or anarthria. Recent advances have demonstrated high-quality text decoding and speech synthesis from electrocorticographic grids placed on the cortical surface.
Here, we investigate a less invasive measurement modality in three participants, namely stereotactic EEG (sEEG) that provides sparse sampling from multiple brain regions, including subcortical regions. To evaluate whether sEEG can also be used to synthesize audio from neural recordings, we employ a recurrent encoder-decoder model based on modern deep learning methods. We find that speech can indeed be reconstructed with correlations up to 0.8 from these minimally invasive recordings, despite limited amounts of training data. In particular, the architecture we employ naturally picks up on the temporal nature of the data and thereby outperforms an existing benchmark based on non-regressive convolutional neural networks. \looseness=-1

\end{abstract}

\section{Introduction}
Brain-Computer Interfaces (BCIs) have progressed tremendously over the last decade and now enable patients who have lost the ability to communicate due to injury, stroke or neuromuscular disorders, to interact with others using a robotic arm  \cite{hochberg2012reach}, high-performance cursor control \cite{pandarinath2017high} or even imagined handwriting \cite{Willett2021}, achieving information transfer rates up to approximately 12 bits/second. Without a doubt, this ability to restore communication through typing or writing improves the quality of life of patients drastically. However, the potential of BCIs does not stop here. In fact, an active line of research is pushing technology for speech decoding from neural signals, as spoken language is still our most natural form of communication with bit rates around 39 bits/second across many languages \cite{coupe2019different}.

Several approaches to directly decode speech from invasive measures of brain activity have been presented in recent years \cite{herff2016automatic,chakrabarti2015progress}.
Martin et al. \cite{10.3389/fneng.2014.00014} decoded spectro-temporal features of speech from electrocorticographic (ECoG) electrodes placed on the cortical surface. In \cite{lotte2015electrocorticographic}, Lotte et al. showed that articulatory features of speech could be decoded from ECoG recordings, which was later investigated in more depth in \cite{chartier2018encoding}. Mugler et al. \cite{mugler2014direct} demonstrated that the full set of American English phonemes can be decoded from ECoG. Combining these decoding successes with approaches from Automatic Speech Recognition (ASR), Herff et al. \cite{herff2015brain} and Moses et al. \cite{moses2016neural} presented that a textual representation of continuous speech could be reliably decoded from ECoG. Moses et al. even demonstrated their approach in real-time \cite{moses2018real,moses2019real}. Decoding a textual representation of speech has the potential to help patients communicate with friends and family. Furthermore, it enables the downstream usage of an ever growing variety of natural language processing tools such as large language models \cite{madotto2020language}. However, crucial semantic information is lost in the textual representation such as intonation, prosody and accentuation. To give patients access to the full expressive power of speech, direct synthesis of speech from neural data is better suited.

Towards this goal, artificial neural networks constitute the most promising models due to their capability of extracting meaningful latent features even when the input space itself has low semantic structure (as is the case for raw EEG signals). Among this class of models, temporal structure and complex auto-correlations of both neural dynamics and speech are best captured using recurrent neural networks that incorporate sequential information through feedback connections. Two recent studies demonstrated that non-recurrent neural networks can also be employed to synthesize produced \cite{angrick2019speech} and perceived \cite{akbari2019towards} speech from ECoG recordings. 
Berezutskaya et al. \cite{berezutskaya2020brain} used recurrent neural networks to predict neural activity from sound features of perceived audio. 
Makin et al. \cite{makin2020machine} employed an encoder-decoder framework to decode a textual representation of speech from ECoG recordings. In a break-through study \cite{moses2021neuroprosthesis}, a closed-loop version of this approach was even used by a patient suffering from anarthria.
Instead of a textual representation, Anumachipalli et al. \cite{Anumanchipalli2019} decoded articulatory gestures from ECoG and translated these gestures into an audio waveform.

All approaches previously discussed utilize brain activity that is measured directly on the cortical surface with ECoG electrodes. These electrodes require a large craniotomy. In the monitoring of epilepsy patients, in which almost all of the previous studies have been conducted, more and more centers utilize the less invasive stereotactic EEG (sEEG), which is measured with intracranial depth electrodes  \cite{van2017methodology}. 
The use of sEEG for BCI has been discussed before \cite{herff2020potential} and successfully applied to word decoding \cite{petrosyan2022speech} and speech synthesis \cite{angrick2022towards}. One study demonstrates that imagined speech can be synthesized in real-time from sEEG recordings
\cite{angrick2021real}. However, the authors opted for a simple decoding pipeline that does not respect the temporal nature of the problem and hence only produces low-quality, unintelligible speech output.

Here, we employ a large neural network architecture that naturally leverage temporal relations to demonstrate the feasibility of speech neuroprostheses based on sEEG by producing audio output. For this purpose, we apply an encoder-decoder architecture with attention mechanisms \cite{vaswani2017attention} to create audible speech directly from neural recordings with intracranial depth electrodes. 

\section{Material \& Methods}
\subsection{Experimental Setup}

\textbf{\large Participants} Three patients (P1 16 y/o male, P2 20 y/o female, P3  40 y/o male) suffering from intractable epilepsy participated in our experiment. All patients were native speakers of Dutch. Patients were implanted with depth electrodes to identify the epileptic foci and plan potential resections. During this mapping procedure, patients participated voluntarily in our experiment and provided written informed consent. The experiment was conducted in accordance with the declaration of Helsinki and approved by the IRB of both Maastricht University and the Epilepsy Center Kempenhaeghe. 

\textbf{\large Experiment and Data Recording} Participants were shown a total of 100 sentences from the Mozilla Common Voice Dutch corpus \cite{ardila2019common} on a computer screen in front of them. All sentences were selected to be between 5 and 7 words long and displayed in pseudo-randomized order. Each sentence was followed by a 2-second rest interval during which a fixation cross was shown on the screen. Duration of each sentence depended on the participants' reading speed leading to total recording lengths between 10 and 20 minutes. We evaluated the proportion of silence in our data by running a very simple Voice-Activity Detection (mean power of the upper half of the frequency ranges with participant-specific thresholds). Audio recordings contained 47.7\%, 47.3\% and 40.8\% of silence for P1, P2 and P3, respectively. This proportion was stable across training and evaluation data.

Neural data were sampled at either 1024 Hz or 2048 Hz using Micromed SD LTM amplifiers (Micromed S.p.A., Treviso, Italy). Electrodes were referenced to a common white matter electrode contact. 
Speech data was recorded using the experiment laptop's built-in microphone and sampled at 48 kHz. Neural data, audio data and the experiment timings were synchronized using LabStreamingLayer \cite{kothe2014lab}. We tested our data for acoustic contamination  using the approach by Roussel et.~al \cite{roussel2020observation}. The risk of falsely rejecting the hypothesis of acoustic contamination is $p<0.01$ for all participants.

Data for this study is available at \url{https://osf.io/7wf6n/}.

\begin{figure}
\centering
\begin{subfigure}{.3\textwidth}
  \centering
  \includegraphics[clip=true,trim=7cm 2cm 7cm 3cm, width=.9\linewidth]{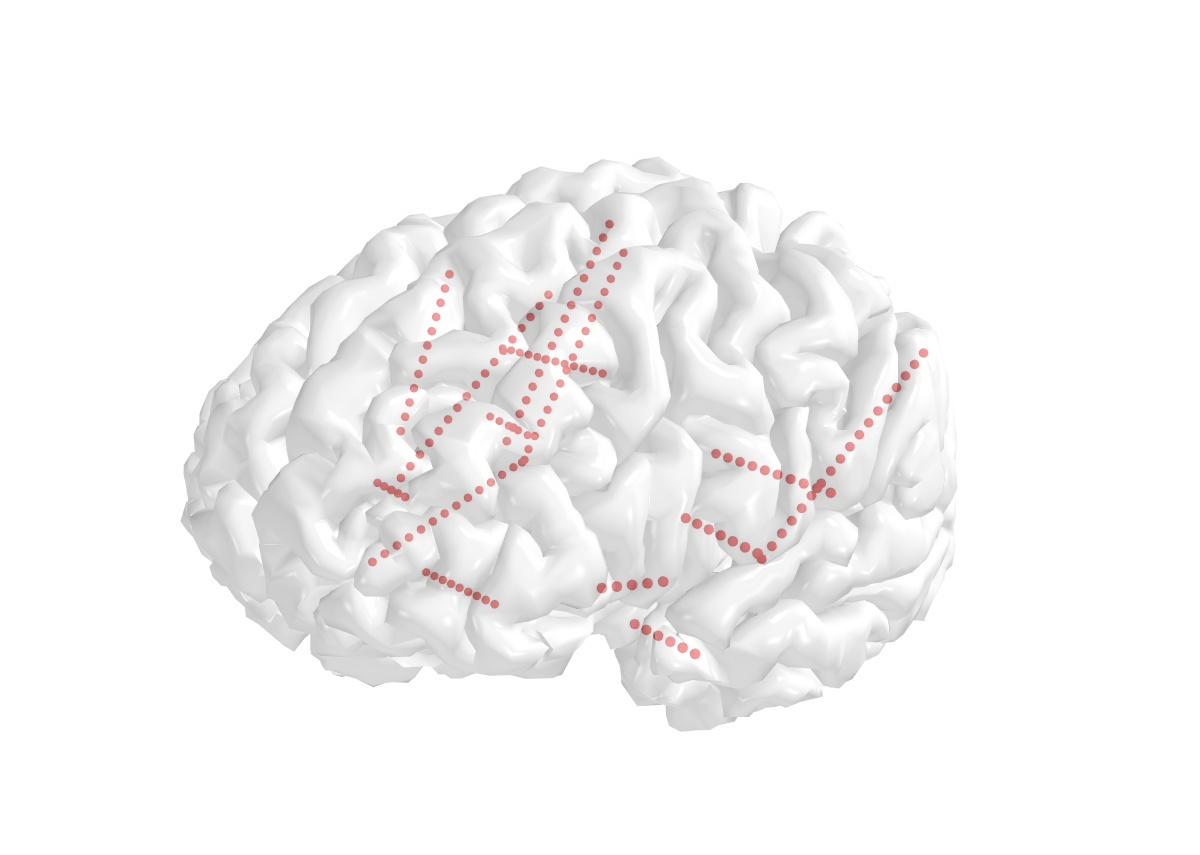}
  \caption{\centering P1}
  \label{fig:sub1}
\end{subfigure}%
\begin{subfigure}{.3\textwidth}
  \centering
  \includegraphics[clip=true,trim=7cm 2cm 7cm 3cm, width=.9\linewidth]{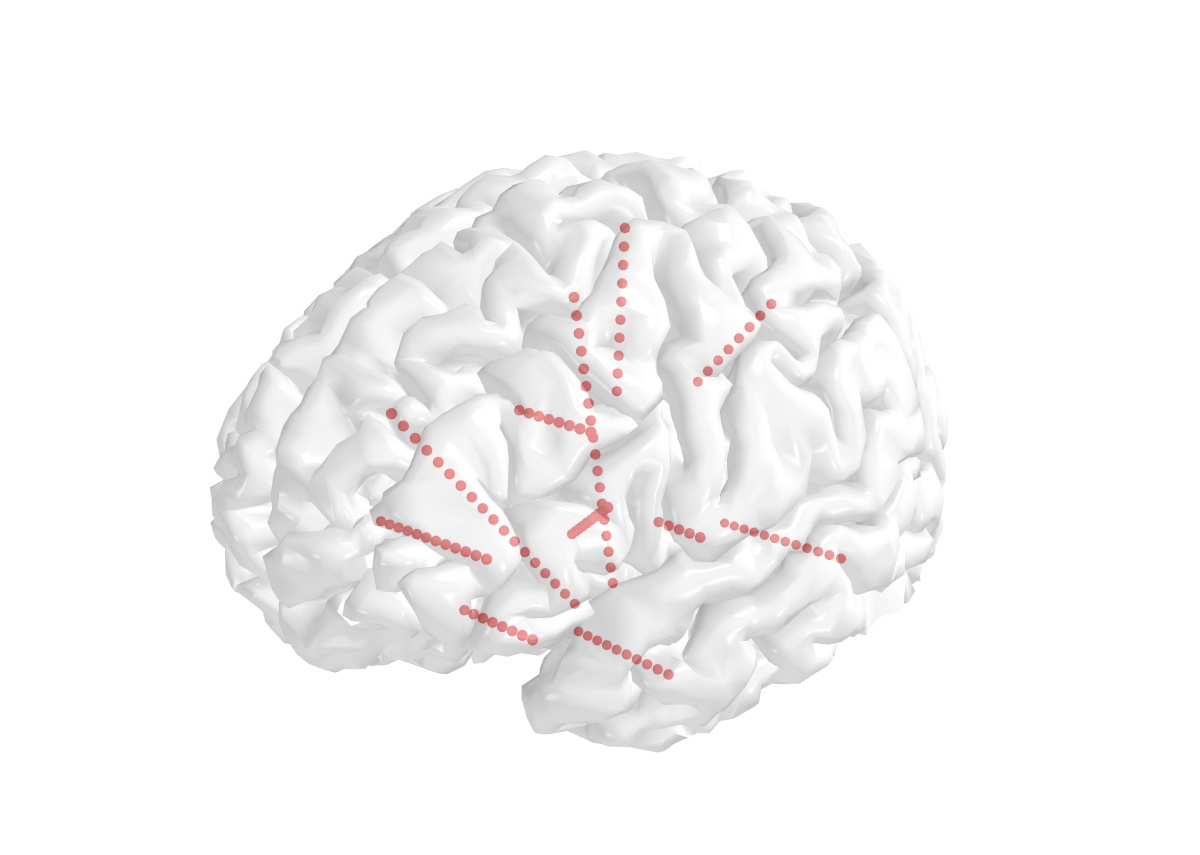}
  
  \caption{\centering P2}
  \label{fig:sub2}
\end{subfigure}
\begin{subfigure}{.3\textwidth}
  \centering
  \includegraphics[clip=true,trim=7cm 2cm 7cm 3cm, width=.9\linewidth]{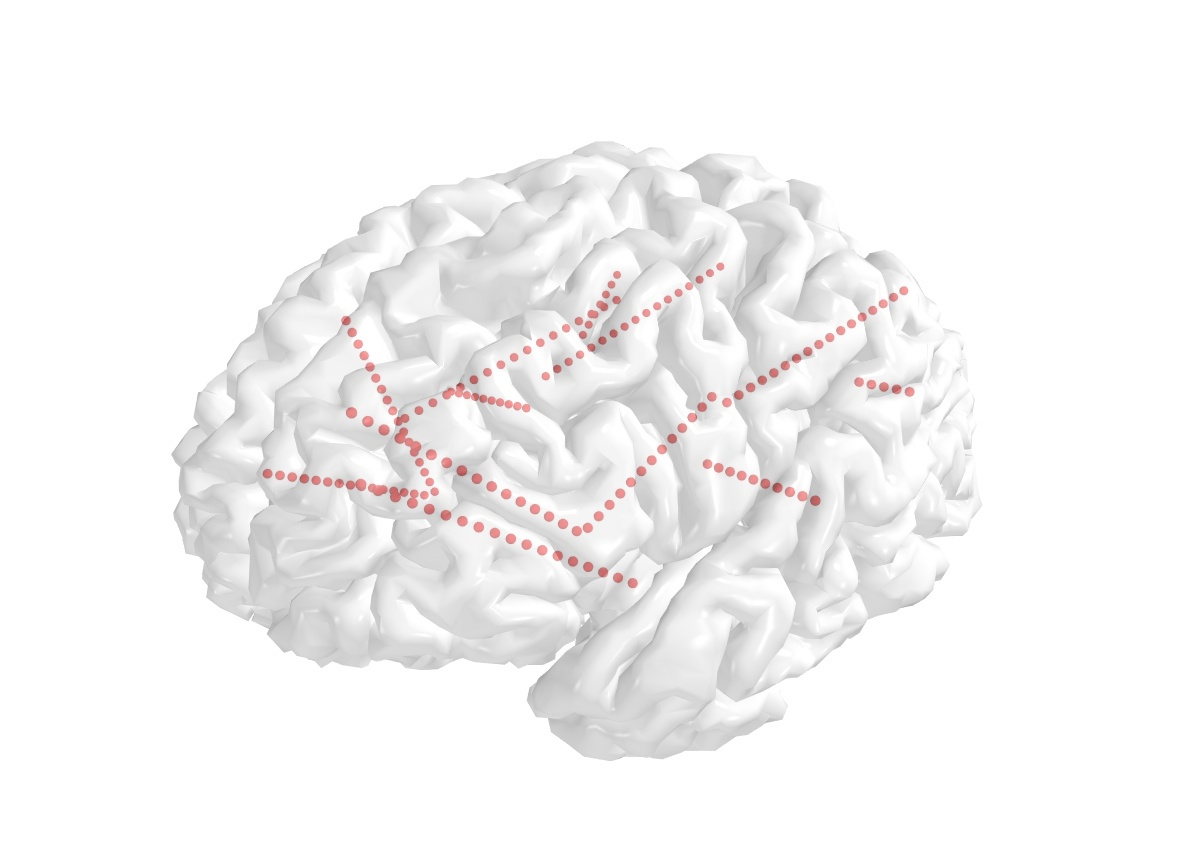}
  \caption{\centering P3}
  \label{fig:sub2}
\end{subfigure}
\caption{Electrode contact locations (red) superimposed on a cortical mesh of the pial surface for all participants. Electrode locations are determined by co-registering pre-surgical MRI and post-surgical CT scans.}
\label{fig:locs}
\end{figure}

\textbf{\large Electrode Localization} Electrode number and locations were purely determined based on clinical necessity. Electrodes with a total of 117, 111 and 127 contacts were implanted for P1, P2 and P3, respectively (Fig.~\ref{fig:locs}). 
Electrode locations were identified using img\_pipe \cite{hamilton2017semi} after co-registering pre-surgical T1-weighted MR scans with post-surgical CT scans. Anatomical labels were acquired from Freesurfer’s \cite{fischl2012freesurfer} cortical parcellation.


\textbf{\large Data Processing}
Speech signals were recorded at $48$ kHz and downsampled to $22'050$ Hz using a pre-computed (kaiser best) filter implemented in LibROSA \cite{mcfee2015librosa}. For each $12.5$ ms block of the recorded speech, we perform a short time Fourier transformation with corresponding re-scaling to obtain a mel-spectrogram representation which is spectrally normalized in a final step using dynamic range compression. All steps follow the procedure from Shen et al. \cite{shen2018natural}, in which $80$ mel frequency coefficients between $0$ and $8000$ Hz are calculated.

We limited our analysis of the sEEG data to the high-gamma band between 70 and 170 Hz, which is routinely employed in studies decoding speech from intracranial recordings \cite{Anumanchipalli2019,akbari2019towards,angrick2021real} as it contains speech \cite{leuthardt2012temporal, crone2001electrocorticographic} and language \cite{towle2008ecog} specific information. This localized information of the high-gamma signal might be explained through the high correlation with ensemble spiking \cite{ray2008neural} and can also be used to identify speech articulatory gestures \cite{chartier2018encoding} and smiling \cite{kern2019human} from intracranial recordings.
The high-gamma band was extracted using an IIR bandpass filter with filter order 8. The first two harmonics of the line noise (50Hz) were attenuated using elliptic IIR notch filters (filter order 8). We then estimated the signal envelope as the magnitude of the analytic signal computed using the Hilbert transform.

\subsection{Decoding Model}
\begin{figure}[h]
    \centering
    \includegraphics[width=0.99\textwidth]{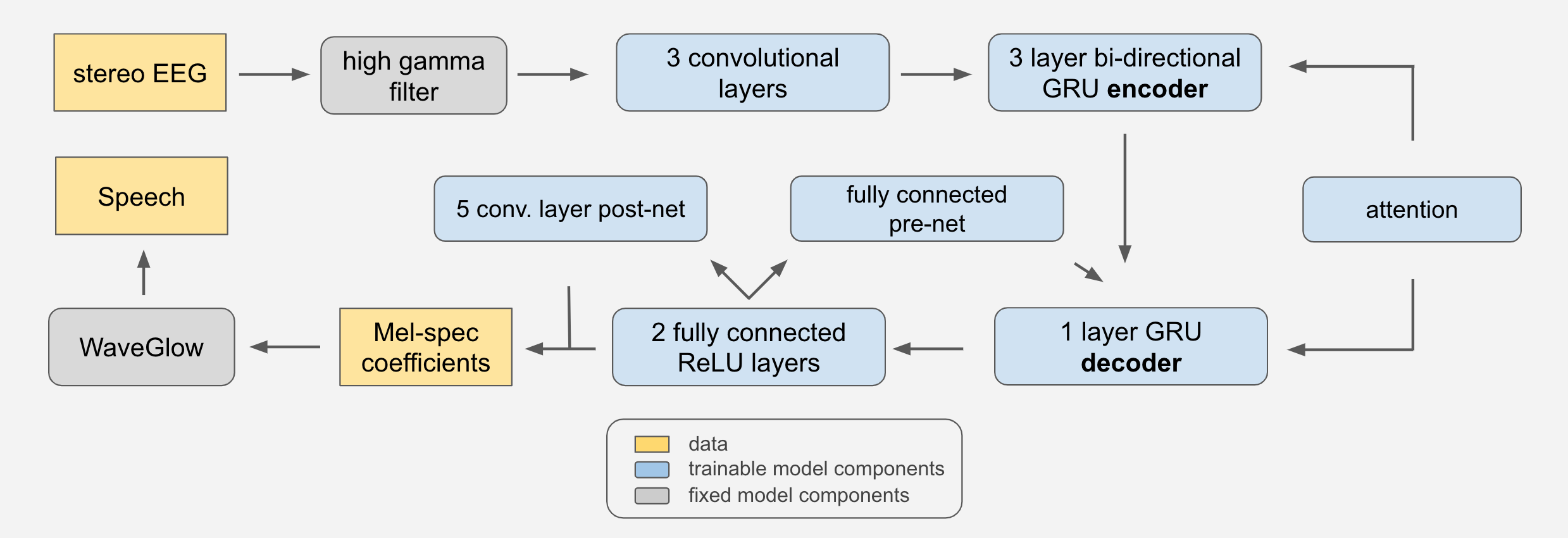}
    \caption{\textbf{Model overview:} Our pipeline consists of two major parts, as indicated by the yellow squares. The first transforms sEEG inputs into time-aligned mel-spectrogram coefficients and is trained end-to-end as detailed out in Section \ref{sec:model_part_i}. The second uses the pre-trained WaveGlow vocoder \cite{prenger2019waveglow} without any modifications to generate speech.\looseness=-1}
    \label{fig:model_overview}
\end{figure}

Our model is composed of a recurrent sequence-to-sequence network, which maps
neural activity to mel-scale spectrograms, and a neural vocoder that synthesizes time-domain
waveforms from the generated spectrograms. While the first component is largely inspired by the Tacotron-2 model \cite{shen2018natural}, which is mainly used in text to speech synthesis tasks, the latter component is a flow based generative model termed WaveGlow \cite{prenger2019waveglow}, which we use without modifications. Figure \ref{fig:model_overview} depicts our model in its entirety. 
Source code for the neural network models as well as some audio samples can be found at \url{https://github.com/jonaskohler/stereoEEG2speech}. Due to limitations in compute power, all hyper-parameters listed in the following sections were kept constant across participants and set according to previous works \cite{shen2018natural,angrick2019speech} for both the model itself as well as for the optimizer.\footnote{Except for the learning rate, which we grid-searched from the candidate set $\{0.1,0.05,0.01,0.005,0.001,0.0005,0.0001\}$ on the validation set of P3.}

\subsubsection{Spectogram regression}\label{sec:model_part_i}
On a high level, the spectogram regression consists of two major steps. We first feed the high gamma band of the neural activity through a sequence of three convolutional layers with one dimensional kernels. These layers act as a filter that computes latent representations of the input sequences, which are then fed into an RNN encoder-decoder based on gated recurrent units (GRUs) \cite{cho2014learning}. This latter network is tasked with mapping the convolved input sequence to a target output sequence of mel spectral coefficients. 

\begin{figure}[h]
    \centering
    \includegraphics[width=0.99\textwidth]{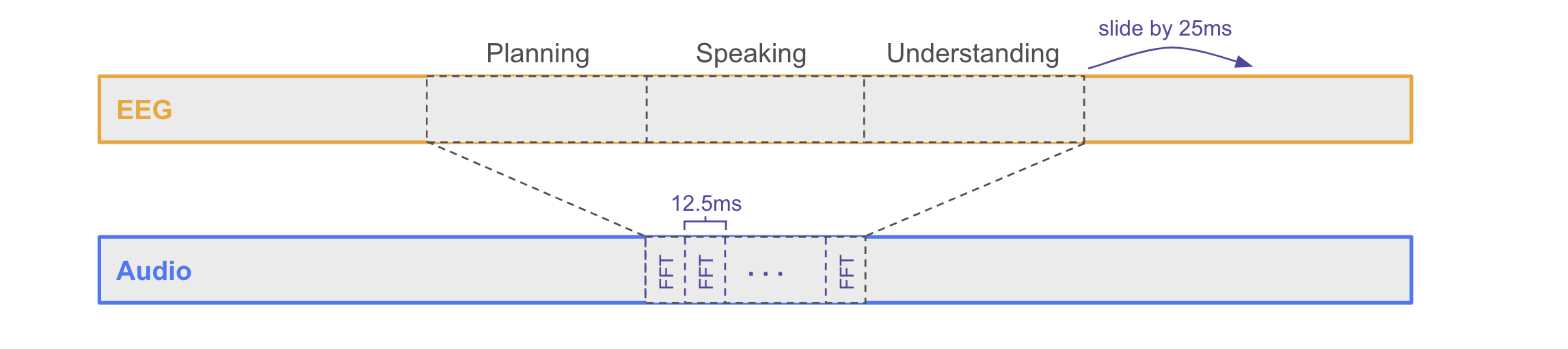}
    \caption{\textbf{Sliding window approach:} To generate input-output samples for the network, we proceed as follows. \textbf{Input:} We process the continuous sEEG sequences in windows of 400ms, which we slide forward with a hop of $25$- (training) and $400$ms (test/validation). For each window, we add an additional $400$ms of sEEG signal before and after the actual speech to account for mental processes like speech planning and understanding. \textbf{Target:} Within each corresponding $400$ms audio window, we perform short time Fourier transformations with corresponding re-scaling of non-overlapping $12.5$ms blocks to obtain a sequence of $32$ mel-spectrogram representation.\looseness=-1}
    \label{fig:sliding_window}
\end{figure}

In this context, the input (EEG) sequence is processed in a sliding window approach. To be precise, we process continuous sequence windows of $400$ms, which we slide through the input with a hop of 25ms (see Fig. \ref{fig:sliding_window})\footnote{For the test and validation sets, the hop is chosen such that the time windows are non-overlapping and continuous.}. For each window, we add an additional $400$ms of sEEG signal before and after the actual speech to account for mental processes like speech planning and understanding. On the target side, we use fast fourier transformations (FFT) to compute spectral representations of speech in 32 non-overlapping blocks ($400 ms / 12.5 ms$) within the 400 ms window.

\textbf{\large CNN} First, the convolutional neural network (CNN) receives sequences of neural activity across all channels at $1024$hz ($~110\times 1024$). This signal is being convolved in three layers with decreasing kernel size and increasing number of channels to result in an output of dimensionality $300$ across $100$Hz. All layers use Batch Normalization \cite{ioffe2015batch} and ReLU activation functions. One dimensional max pooling is applied in the last layer. \\

\textbf{\large RNN} Secondly, we add positional embeddings to the convolved input sequence \cite{vaswani2017attention}, reverse its time order and then feed it into a three-layer bi-directional RNN encoder with gated recurrent units \cite{cho2014learning}. The encoder converts this sequence into a hidden feature representation of dimensionality $333$, which the decoder consumes to predict a spectrogram in an autoregressive manner (i.e. one step at a time). In this typical enconder-decoder setting, the information of the entire input sequence is compressed into a \textit{single} fixed-length vector (i.e. the hidden state of the last time step), which makes it hard for the decoder to cope with long input sequences. As our sEEG input sequences are much longer than for example sentences in natural language, we employ an additive attention mechanism termed, Bahdnau-Attention \cite{bahdanau2014neural}, which allows the decoder to furthermore incorporate information from the hidden states of \textit{any} time step of the input sequence.

Inspired by Tacotron-2, the decoder output is post-processed by a set of two linear layers (pre-net), which act as an information bottleneck, before it is fed back into the decoder as initial state for the next timestep. Furthermore, while the decoder output is concatenated to the attention context vector and projected through a linear transform to predict the target spectrogram frame as usual, it is also passed through a 5-layer convolutional post-net which predicts a residual to add to the prediction to improve the overall reconstruction.\\

\textbf{\large Training} We train the feature extractor (CNN) and spectogram predictor (RNN) in an end-to-end fashion, applying the standard maximum-likelihood training procedure with a mean-squared-error loss and a teacher forcing ratio of $0.1$. That is, in $10\%$ of the cases we replace the decoder prediction from the previous state with the ground truth. We employ the AdamW optimizer with default parameters \cite{loshchilov2017decoupled}, batch-size $512$ and learning rate $\eta=0.0005$ as well as a weight decay of $\lambda=0.001$. We train for a fixed number of $50$ epochs and employ a learning rate scheduler which decreases $\eta$ by a factor of two in epoch 45. Our architecture has a total of 10'411'964 trainable parameters.\\

To be precise, our model transforms a sequence of sEEG input with $c$ channels $\bm{X} \in \mathbb{R}^{c \times T_1}$ into a sequence of $80$ dimensional mel spectrogram coefficients $\bm{\hat{Y}} \in \mathbb{R}^{80 \times T_2}$ and minimizes

\begin{equation}\label{eq:loss}
    \mathcal{L(\mathbf{W})}= \frac{1}{2} \|\bm{Y}-\bm{\hat{Y}}\|_F^2, \quad \bm{\hat{Y}}:= f_{NN}\left(\bm{X},\bm{W}\right),
\end{equation}
where $\bm{\bm{Y}}\in \mathbb{R}^{80 \times T_2}$ contains the ground truth mel-coefficients and $f_{NN}:\mathbb{R}^{c \times T_1}\rightarrow \mathbb{R}^{80\times T_2} $ represents the network mapping which is parametrized by a set of weights $\bm{W}$. The specific values $T_1$ and $T_2$ are determined by the window- and hop size with which we move through the data ($T_1$) as well as by the interval in which we compute mel spectograms ($T_2$). See Fig. \ref{fig:sliding_window}) for an illustration. \\

\textbf{\large Baseline Comparison} To establish whether the proposed model outperforms previously described approaches, we compare our results to a synthesis approach based on a densely connected convolutional neural network (DenseNet), as presented in \cite{angrick2019speech}. Notably, this approach significantly outperforms a chance baseline designed by the authors for the specific task at hand. We follow exactly the implementations detailed there. The architecture is composed of densely connected convolutional blocks and transition blocks in an alternating manner (see Fig. 3 in the original paper). Contrary to our encoder-decoder network, we feed windows of $50$ms length into this network as this is what the network architecture was optimized for in the original work. Also along the lines of \cite{angrick2019speech}, we split the channel dimension into two to generate (preferably) square two dimensional inputs for each time step (network employs three dimensional convolutional kernels). To ensure comparability between the two approaches, we feed in raw time series, as we do for the recurrent architecture, instead of averages in windows as utilized in the original study. While this differs from the original implementation, it allows for more meaningful comparison between the two architectures. We evaluated the DenseNet output to ensure that this change in pre-processing still lead to meaningful results.
Our implementation of the baseline system has 86'020 trainable parameters.

\subsubsection{Speech generation}
After completing the first step of transforming sEEG signal to mel-spectrograms, we employ a flow-based generative model to turn these time-aligned features into an audio waveform. Specifically, we employ Nvidia's WaveGlow model \cite{prenger2019waveglow}, which is designed to provide fast, efficient and high-quality audio synthesis without the need for auto-regression, as a plug and play decoder to synthesize speech. We use this synthesis approach for both the proposed encoder-decoder architecture as well as the baseline from \cite{angrick2019speech}.

\subsection{Intelligibility test}
To evaluate a first indicator of the intelligibility of the produced audio, we conducted a forced-choice intelligibility test in which 19 healthy volunteers listened to all reconstructed sentences from the test set one by one and had to select the textual representation they thought they had heard out of 2 possible text options.
For this purpose, the actual text prompt and a random text prompt from one of the other sentences were offered as choices.
As all sentences in the data set have between 5 and 7 words, the sentences are somewhat balanced for length. The intelligibility test was implemented in BeagleJS \cite{kraft2014beaqlejs}.

\section{Results}

\subsection{Speech can be generated from intracranial depth electrodes}
We employed a fixed training-validation-test split, where both the validation and test set have a fixed length of 1.5 minutes for each set and participant. The remaining data were used as training set, whose specific length per participant is given in the first column of Table \ref{tbl:loss}.


\begin{figure}

\begin{tabular}{c@{\hspace{0.001pt}}c@{\hspace{0.001pt}}c@{\hspace{5pt}}c@{\hspace{0.001pt}}}
&\small \textit{"Het restaurant ligt op de Oratoriënberg."}& & \small \textit{"Het restaurant ligt op de Oratoriënberg."}\\
\rotatebox{90}{\scriptsize \hspace{-20pt}Magnitude} &
\centering
\begin{subfigure}{.47\textwidth}
  \centering
  \includegraphics[width=0.99\linewidth,trim=20 20 20 20,clip]{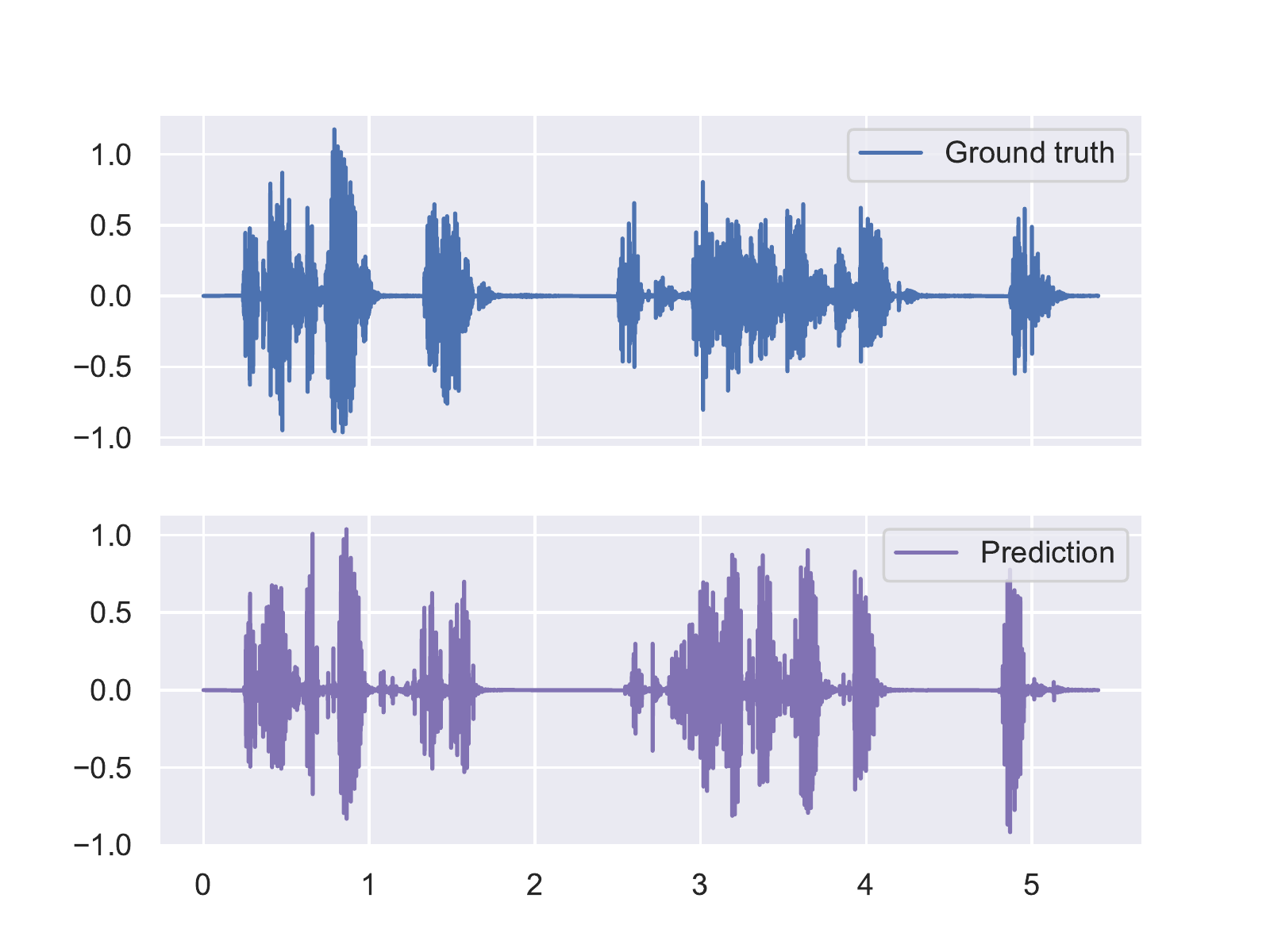}
\end{subfigure}&
\rotatebox{90}{\scriptsize \hspace{-20pt} Frequency} 
&
\begin{subfigure}{.47\textwidth}
  \centering
  \includegraphics[width=0.99\linewidth,trim=20 20 20 20,clip]{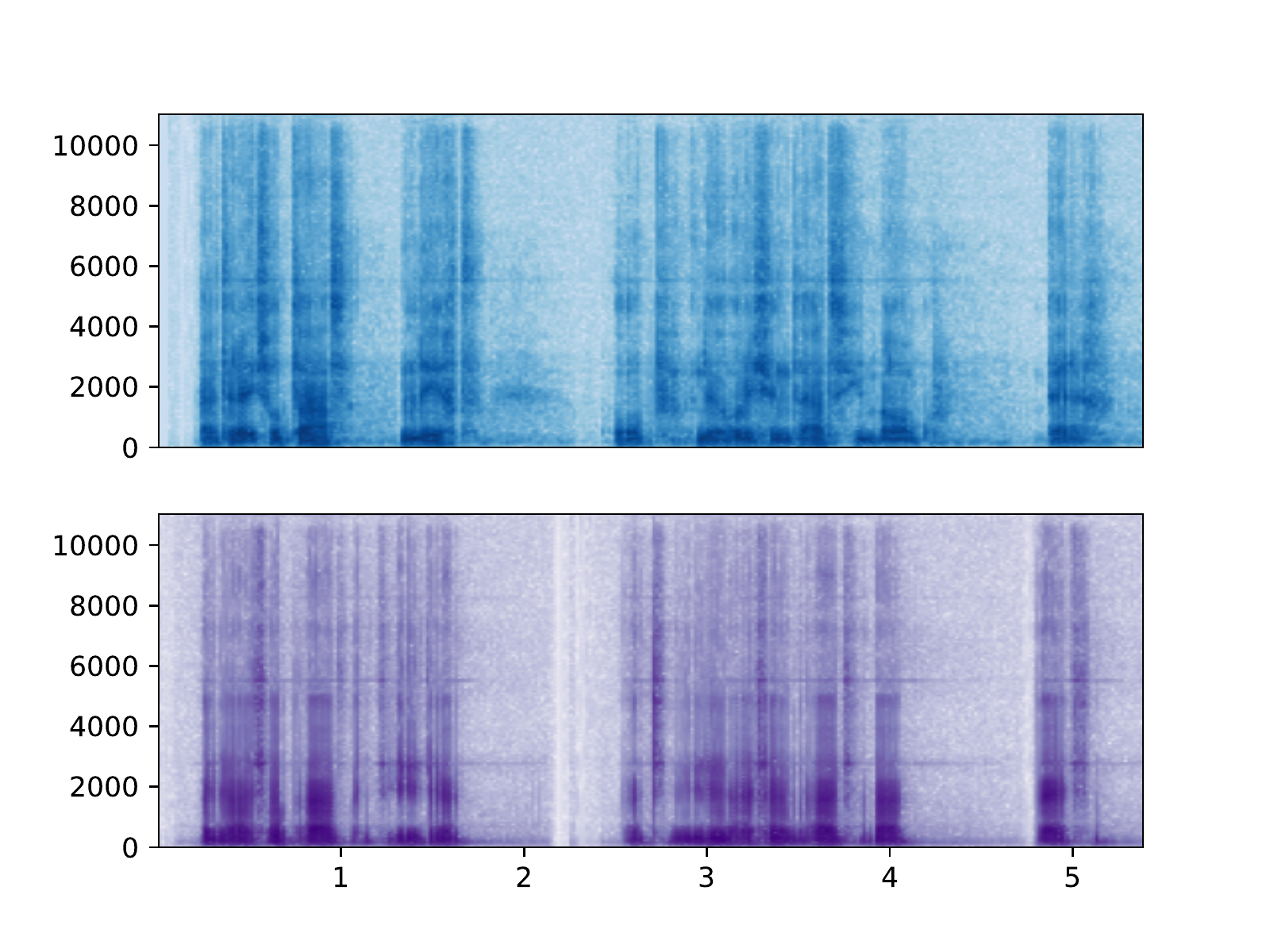}
\end{subfigure}\\
  &\scriptsize Seconds&&\scriptsize Seconds\end{tabular}
\caption{An exemplary sentence (5.5 seconds) from the test set of P3 in waveform (left) and spectogram (right) representation. }
\label{fig:example_sentence}
\end{figure}

Our method produces audio waveforms that appear similar to the ground truth to the human eye (see Figure \ref{fig:example_sentence} for a particularly successful sentence of P3).
When first primed with the ground truth, the generated audio is sometimes even comprehensible. Furthermore, our combination of the encoder-decoder framework with the WaveGlow vocoder succeeds in preserving voice characteristics to some degree.\footnote{Audio samples that account for these two claims can be found here: \url{https://github.com/jonaskohler/stereoEEG2speech}} These results demonstrate that intracranial depth electrodes can be used to synthesize audio, despite the suboptimal sampling across many brain regions instead of the focused sampling of relevant areas provided by ECoG. Notably, $400$ms of sEEG are processed by our model in roughly $80$ms on a single NVIDIA P100 GPU.

\begin{table}[h]
\centering
\begin{tabular}{|l|l|l|ll|}
\hline
\headrow
          & Training & Validation & \multicolumn{2}{c}{Test} \\
          & Ours         & Ours               & Ours   & DenseNet      \\
          \hline
          
          P1 (5min) & 0.24   $\pm 0.02$             &     1.9    $\pm 0.2$      &   \bf{ 2.1$\bm{\pm 0.1}$       }    &   7.4 $\pm 2.0$  \\
P2 (14min) &    0.4 $\pm 0.01$       &           1.5  $\pm 0.2$    &      \bf{ 1.8 $\bm{\pm 0.1}$}     &  10.3 $\pm 3.1$     \\
P3 (17min) &    0.31   $\pm 0.02$         &      1.3  $\pm 0.1$               &      \bf{ 1.4 $\bm{\pm 0.2}$}             & 7.1 $\pm 2.7$  \\
 \hline   
\end{tabular}

\caption{Mean-squared error loss (on mel spectral coefficients) of our model on training, validation and test set. As a baseline, we also report numbers for the DenseNet from \cite{angrick2019speech}, which in turn beat a sophisticated randomized baseline (see Figure 5 there). Mean ($\pm$ standard deviation) of five independent random initializations.}
\label{tbl:loss}
\end{table}

\begin{figure}
\centering
\begin{subfigure}{1.\textwidth}
  \centering
  \includegraphics[width=.66\linewidth,trim=5 8 5 22,clip]{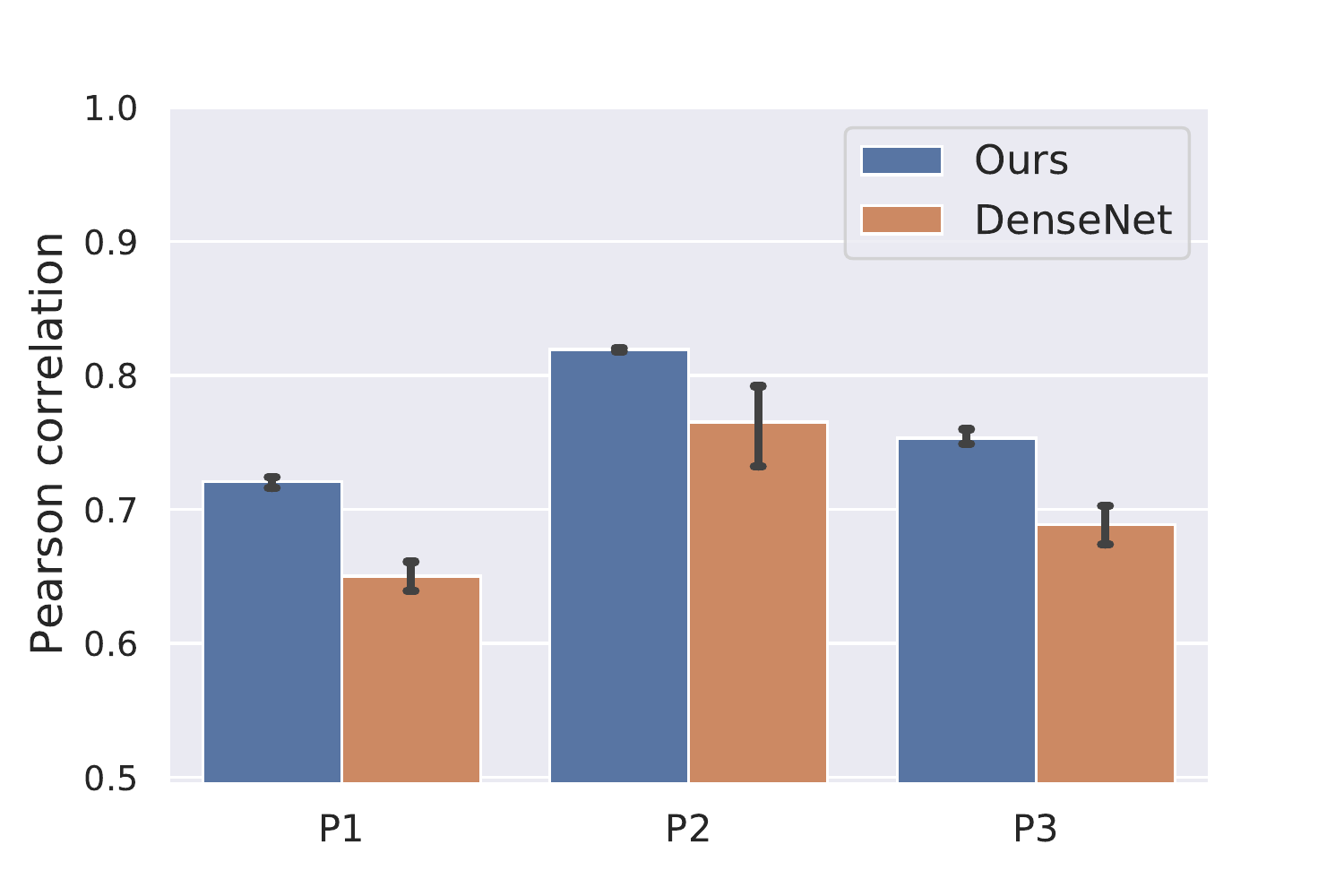}
  \caption{\centering}
  \label{fig:test_sub1}
\end{subfigure}%

\begin{subfigure}{.45\textwidth}
  \centering
  \includegraphics[width=1\linewidth,trim=5 5 5 5,clip]{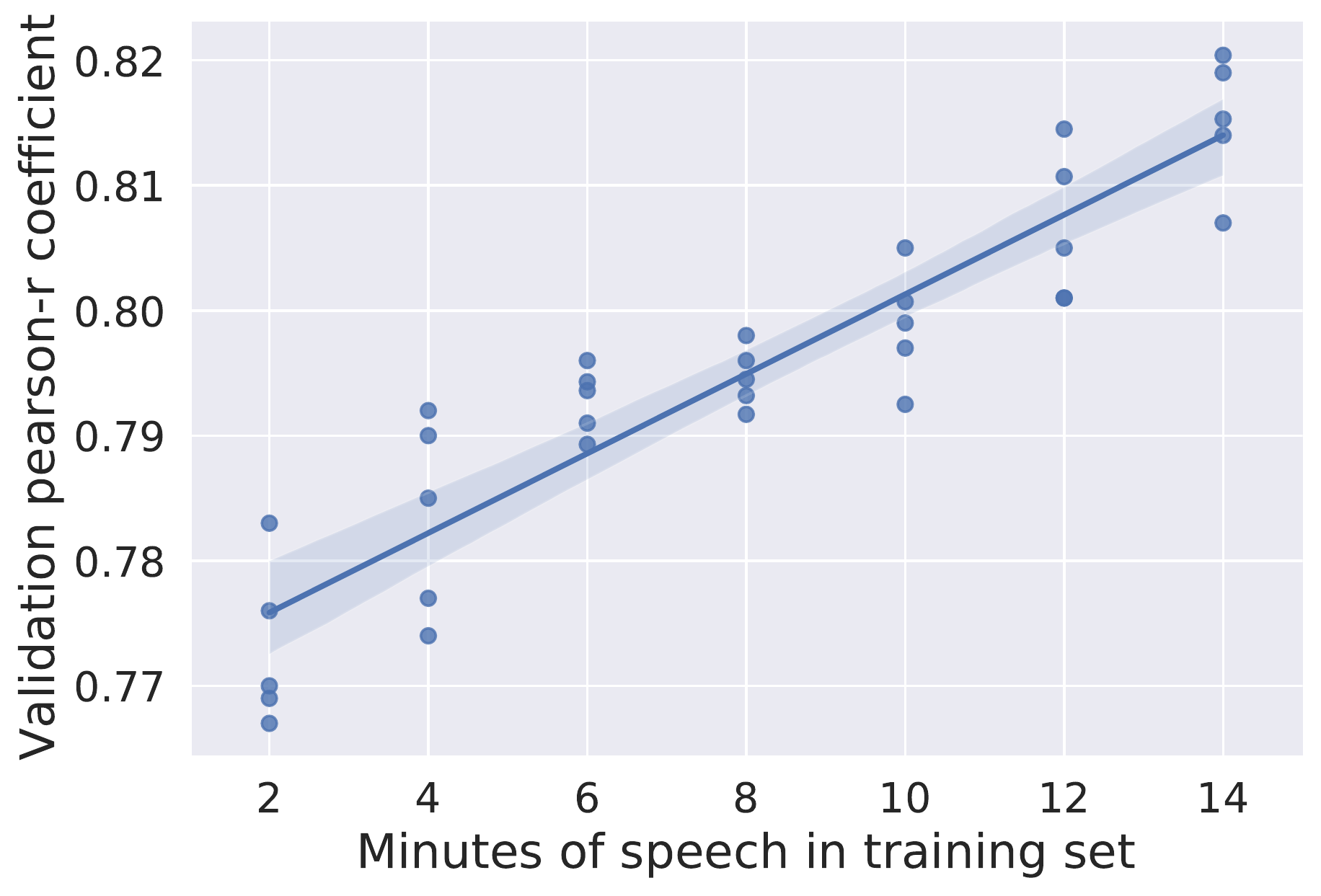}
  \caption{\centering}
  \label{fig:test_sub2}
\end{subfigure}
\begin{subfigure}{.45\textwidth}
  \centering
  \includegraphics[width=.9\linewidth,trim=7 7 7 14,clip]{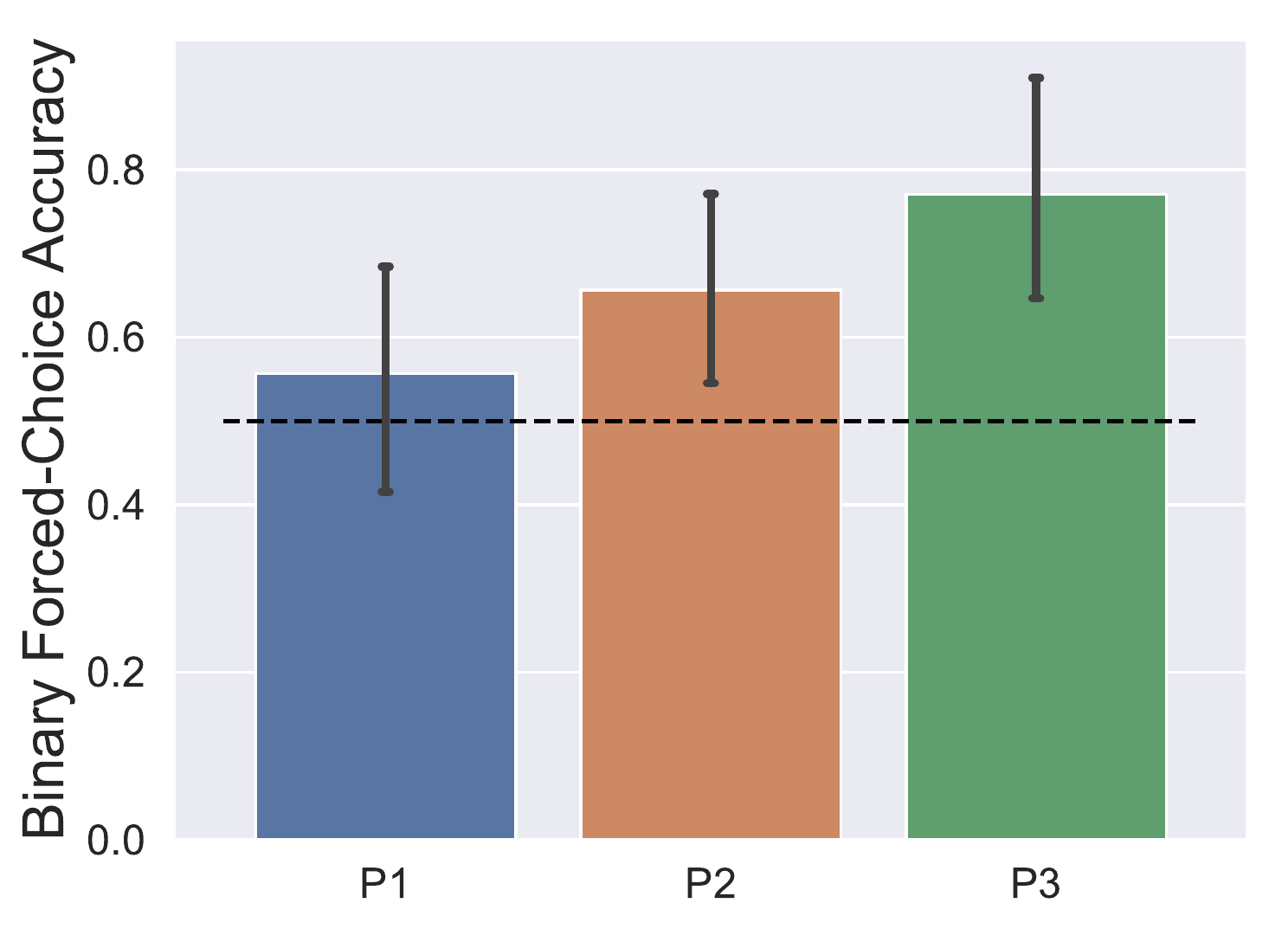}
  
  \caption{\centering}
  \label{fig:test_sub2}
\end{subfigure}
\caption{Decoding Results on the test set. (A) Pearson correlation coefficient for different settings and participants. Mean and 95\% confidence interval of 5 independent runs. The proposed methods outperforms the DenseNet baseline for all three participants. (B) Correlation coefficients for P2 on training sets of increasing length. Five independent runs for each setting. Results still increase with large training set sizes suggesting that improved results can be expected when more data is available ($r=0.92, p<0.001$). (C) Accuracies in binary forced-choice listening test. Sentences of P2 and P3 are discriminable significantly above chance level (dashed line). }
\label{fig:results}
\end{figure}

We evaluate our results using the mean-squared-error loss and the Pearson correlation between reconstructed and original spectrograms. For the Pearson correlation, each mel-scaled spectral bin is correlated over time individually and then the average is taken. While the Pearson correlation is not a perfect measure of speech quality, it was recently shown to better correlate with intelligibility then other measures \cite{berezutskaya2022direct}. 
In comparison to existing works, our approach outperforms the baseline from \cite{angrick2019speech} both in terms of mean-squared-error loss (Tab. \ref{tbl:loss}) and Pearson correlation coefficient (Fig. \ref{fig:results} (a)) significantly (t-tests, $p<0.05$). Note that we only measure performance against the previously presented DenseNet architecture \cite{angrick2019speech} and not against a random baseline, as the DenseNet greatly outperformed a chance level.
The better results might be explained by the fact that the proposed recurrent encoder-decoder architecture is able to explicitly model the temporal nature of both the neural- as well as the audio data and is thereby better able to capture the intricate temporal dynamics of both timeseries.

It is important to note the very limited amount of training data in this setting. As can be seen in Table \ref{tbl:loss}, the network runs the risk of overfitting the few training sentences it has seen from P1 (notice how P1 shows lowest training- but highest test error). Even the longest recording (P3) only contains 17 minutes of training data, of which more than half is silence. We thus explored the influence of training set size on reconstruction quality (Fig. \ref{fig:results} (b)) and observed a clear improvement with more training data. Minutes in the training data and corresponding Pearson correlations on the validation set correlated strongly ($r=0.92$). This analysis was only carried out for P3, as most data was available.
Given the large amount of trainable parameters, it is not surprising that our architecture can still benefit from more training data.
The fact that reconstruction quality does not plateau gives rise to the hope that improved results can be expected with more training data. 

To evaluate whether the reconstructed audio could be helpful for communication, we conducted a forced-choice listening test, in which healthy volunteers had to decide which one of two textual options a reconstructed audio was. For two out of the three participants, this resulted in above chance level identification of the correct sentence (Fig. \ref{fig:results} (c)). For statistical comparison, we generated a set of as many random answers as we had participants in our listening test. This procedure was repeated 10,000 times to generate a distribution of random results. We then drew random results from this distribution for each sentence of a participant and averaged the resulting accuracy. This procedure was again repeated 10,000 and we then looked at the 5\% best accuracies for each participant as an upper limit for chance accuracies. Listening test accuracies for P1 were below this threshold, while results for P2 and P3 are significantly higher than 95\% of random accuracies.
For P1 with the lowest Pearson correlation results, the reconstructed sentences were also not identifiable.

\subsection{Temporal Context is similar in audio and neural data}

\begin{figure}[h]

\begin{tabular}{c@{\hspace{0.001pt}}c@{\hspace{0.001pt}}c@{\hspace{0.001pt}}c@{\hspace{0.001pt}}}

\rotatebox{90}{\scriptsize \hspace{-20pt}Decoder timesteps (ms)} &
\centering
\begin{subfigure}{.33\textwidth}
  \centering
  \includegraphics[width=0.99\linewidth,trim=20 20 20 20,clip]{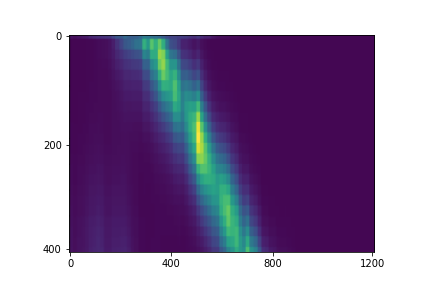}
\end{subfigure}&
\begin{subfigure}{.33\textwidth}
  \centering
  \includegraphics[width=0.99\linewidth,trim=20 20 20 20,clip]{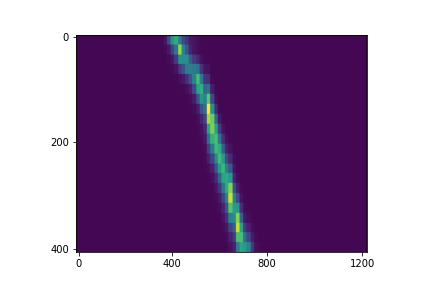}
\end{subfigure}&
\begin{subfigure}{.33\textwidth}
  \centering
  \includegraphics[width=0.99\linewidth,trim=20 20 20 20,clip]{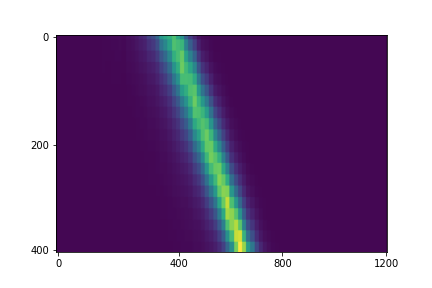}
\end{subfigure}\\
  &&\scriptsize{Encoder timesteps (ms)}&\\
  
 & (a) P1& (b) P2& (c) P3
  
\end{tabular}
\caption{Exemplary attention matrix visualizing attention scores (from randomly selected training samples) at convergence for the three participants (brighter values indicate higher attention scores). Time steps in the decoder are depicted on the y-axis, encoder timesteps are depicted on the x-axis. The diagonal structure suggests that the attention scores are well aligned in the time domain, as for example latter steps in the output attend to latter steps in the input. The figure is furthermore suggestive of the fact that padding the input sEEG sequence (speech planning and understanding) might be unnecessary, as not much attention is paid to the very first and very last input steps. }
\label{fig:attention}
\end{figure}

Fig. \ref{fig:attention} visualizes attention matrices which illustrate  the temporal context in the encoder (x-axis) that the decoder (y-axis) attends to. This temporal context appears to be well aligned between neural data and corresponding audio, which is shown by the diagonal structure of the high attention scores (bright colors). Interestingly,  attentions scores are comparatively low in the early and late parts of the encoder sequence, suggesting that the padding of additional 400 ms before and after the targeted speech segment might not be necessary. This is somewhat contradictory to prior investigation in the temporal context of speech production \cite{brumberg2016spatio}, but might be explained by the recurrent nature of our model.As the sequence to sequence model incorporates temporal structure of the data, smaller context might be necessary to include the complete information. 

As a small ablation study, we carried out ten additional training runs for padding windows of $0$ms and $200$ms on Patient 3. While no padding at all yielded an average test MSE of $1.61 \pm 0.049$, the model achieved a $1.44 \pm 0.06$ MSE when given $200$ms padding. Comparing this with the original MSE of $1.42 \pm 0.054$ (see table \ref{tbl:loss}), we find that zero padding can in fact be considered significantly worse ($p=0.0003$, paired t-test). However, our small study suggests that recurrent models can faithfully be run with just $50\%$ of sEEG padding for speech planning and understanding (t-test found no significant difference in the MSE between the $200$ and $400$ms runs, $p=0.5284$).

\section{Discussion}

In this study, we demonstrated that minimally invasive recordings of neural activity can be used to synthesize audio using a encoder-decoder framework. The similarity of the procedure to implantation of deep brain stimulation electrodes, which are routinely implanted for many years, gives hope for the feasibility in patient cohorts \cite{herff2020potential}. 
Despite the suboptimal sampling of distributed brain regions, the sampled regions provide enough information for speech reconstruction. This is explainable by the large amount of regions involved in speech perception \cite{hickok2000towards} and production \cite{hickok2012computational}. Recent findings identify more and more brain areas involved in these intricate processes, including for example the hippocampus \cite{van2020hippocampus}. 

Quantitative results for the proposed method outperform a previously presented network both in terms of Mean-Squared Error, as well as in terms of Pearson Correlation. While the proposed method is better in both measures, the difference between both approaches is much larger in Mean-Squared Error then in the Pearson correlation. This could be explained by the implicit scaling done by correlation measures. Importantly, this again demonstrates the need for better metrics to evaluate the current level of speech synthesis from neural recordings. 

Despite the very promising results achieved with our approach, it is important to note that all results were produced on previously recorded, offline data of patients that were speaking audibly. Furthermore, the long temporal context used in our approach combined with the long processing time of our encoder-decoder framework prevent our approach from being applicable to a real-time scenario. With this, the approach has some of the inherent disadvantages of approaches that decode a textual representation of speech \cite{moses2016neural,makin2020machine,herff2015brain}, namely that they cannot provide the natural flow of a conversation. To enable this, real-time synthesis of neural data is necessary \cite{angrick2021real}. Our results on the temporal context of the architecture point out that shorter temporal contexts might suffice, bringing us closer to real-time readiness. However, even the 400 ms delay introduced by our current audio framesize would have negative impact on natural speech processes \cite{stuart2002effect}.

The employed WaveGlow architecture for reconstruction of audio waveforms from spectral representation is already real-time ready. Our approach does capture the participants' own voice and is potentially capable of reconstructing speech information beyond the words, such as prosody and accentuation.

In the data presented here, participants spoke naturally. For a speech neuroprosthesis to be useful to patients, it needs to function on imagined or attempted speech processes. Two studies have investigated decoding textual representations from attempted speech \cite{moses2021neuroprosthesis} and speech synthesis from imagined speech \cite{angrick2021real}. Investigating attempted or imagined speech processes without immediate feedback is challenging, so it is outside of the scope of this first feasibility study for reconstruction from sEEG.

Compared to standard text-to-speech (TTS) approaches, our approach is trained on tiny amounts of relatively noisy data. Our analysis (Fig. \ref{fig:results} (b)) highlight that results have not saturated yet, and that more training data is still expected to improve reconstruction results. Alternatively, techniques such as data augmentation and ensembling should improve performance.
Additionally, fine tuning the speech decoder on our dataset is likely to further improve audio quality, which is a relevant step once speech BCIs approach clinical application.

Despite these limitations, our study demonstrates the large potential of encoder-decoder based deep learning models to produce speech reconstruction from minimally invasive neural recordings.

\section{Data and Code Availability}
Code used in this study is available on \url{https://github.com/jonaskohler/stereoEEG2speech}. All data used in this study is available on \url{https://osf.io/7wf6n/}. Note that the participants' voices are anonymized.

\section*{Acknowledgements}
We are grateful to Yannic Kilcher for insightful discussions.
C.H. acknowledges funding by the Dutch Research Council (NWO)  through  the  research  project  ’Decoding  Speech  In SEEG (DESIS)’ with project number VI.Veni.194.021.

\printendnotes

\bibliography{sample}

\begin{thebibliography}{51}
\providecommand{\natexlab}[1]{#1}
\providecommand{\url}[1]{\texttt{#1}}
\providecommand{\urlprefix}{}

\bibitem[{Hochberg et~al.(2012)Hochberg, Leigh R and Bacher, Daniel and
  Jarosiewicz, Beata and Masse, Nicolas Y and Simeral, John D and Vogel, Joern
  and Haddadin, Sami and Liu, Jie and Cash, Sydney S and Van Der Smagt, Patrick
  and others}]{hochberg2012reach}
Hochberg LR, Bacher D, Jarosiewicz B, Masse NY, Simeral JD, Vogel J, et~al.
\newblock Reach and grasp by people with tetraplegia using a neurally
  controlled robotic arm.
\newblock Nature 2012;485(7398):372--375.

\bibitem[{Pandarinath et~al.(2017)Pandarinath, Chethan and Nuyujukian, Paul and
  Blabe, Christine H and Sorice, Brittany L and Saab, Jad and Willett, Francis
  R and Hochberg, Leigh R and Shenoy, Krishna V and Henderson, Jaimie
  M}]{pandarinath2017high}
Pandarinath C, Nuyujukian P, Blabe CH, Sorice BL, Saab J, Willett FR, et~al.
\newblock High performance communication by people with paralysis using an
  intracortical brain-computer interface.
\newblock Elife 2017;6:e18554.

\bibitem[{Willett et~al.(2021)Willett, Francis R. and Avansino, Donald T. and
  Hochberg, Leigh R. and Henderson, Jaimie M. and Shenoy, Krishna
  V.}]{Willett2021}
Willett FR, Avansino DT, Hochberg LR, Henderson JM, Shenoy KV.
\newblock High-performance brain-to-text communication via handwriting.
\newblock Nature 2021 May;593(7858):249--254.
\newblock \urlprefix\url{https://doi.org/10.1038/s41586-021-03506-2}.

\bibitem[{Coup{\'e} et~al.(2019)Coup{\'e}, Christophe and Oh, Yoon Mi and
  Dediu, Dan and Pellegrino, Fran{\c{c}}ois}]{coupe2019different}
Coup{\'e} C, Oh YM, Dediu D, Pellegrino F.
\newblock Different languages, similar encoding efficiency: Comparable
  information rates across the human communicative niche.
\newblock Science advances 2019;5(9):eaaw2594.

\bibitem[{Herff and Schultz(2016)Herff, C. and Schultz,
  T.}]{herff2016automatic}
Herff C, Schultz T.
\newblock {Automatic speech recognition from neural signals: a focused review}.
\newblock Frontiers in neuroscience 2016;10.

\bibitem[{Chakrabarti et~al.(2015)Chakrabarti, Shreya and Sandberg, Hilary M
  and Brumberg, Jonathan S and Krusienski, Dean J}]{chakrabarti2015progress}
Chakrabarti S, Sandberg HM, Brumberg JS, Krusienski DJ.
\newblock Progress in speech decoding from the electrocorticogram.
\newblock Biomedical Engineering Letters 2015;5(1):10--21.

\bibitem[{Martin et~al.(2014)Martin, S. and Brunner, P. and Holdgraf, C. and
  Heinze, H.-J. and Crone, N.E. and Rieger, J. and Schalk, G. and Knight, R.T.
  and Pasley, B.}]{10.3389/fneng.2014.00014}
Martin S, Brunner P, Holdgraf C, Heinze HJ, Crone NE, Rieger J, et~al.
\newblock {Decoding spectrotemporal features of overt and covert speech from
  the human cortex}.
\newblock Frontiers in Neuroengineering 2014;7(14).

\bibitem[{Lotte et~al.(2015)Lotte, Fabien and Brumberg, Jonathan S and Brunner,
  Peter and Gunduz, Aysegul and Ritaccio, Anthony L and Guan, Cuntai and
  Schalk, Gerwin}]{lotte2015electrocorticographic}
Lotte F, Brumberg JS, Brunner P, Gunduz A, Ritaccio AL, Guan C, et~al.
\newblock Electrocorticographic representations of segmental features in
  continuous speech.
\newblock Frontiers in human neuroscience 2015;9.

\bibitem[{Chartier et~al.(2018)Chartier, Josh and Anumanchipalli, Gopala K and
  Johnson, Keith and Chang, Edward F}]{chartier2018encoding}
Chartier J, Anumanchipalli GK, Johnson K, Chang EF.
\newblock Encoding of articulatory kinematic trajectories in human speech
  sensorimotor cortex.
\newblock Neuron 2018;98(5):1042--1054.

\bibitem[{Mugler et~al.(2014)Mugler, Emily M and Patton, James L and Flint,
  Robert D and Wright, Zachary A and Schuele, Stephan U and Rosenow, Joshua and
  Shih, Jerry J and Krusienski, Dean J and Slutzky, Marc W}]{mugler2014direct}
Mugler EM, Patton JL, Flint RD, Wright ZA, Schuele SU, Rosenow J, et~al.
\newblock Direct classification of all American English phonemes using signals
  from functional speech motor cortex.
\newblock Journal of neural engineering 2014;11(3):035015.

\bibitem[{Herff et~al.(2015)Herff, Christian and Heger, Dominic and De Pesters,
  Adriana and Telaar, Dominic and Brunner, Peter and Schalk, Gerwin and
  Schultz, Tanja}]{herff2015brain}
Herff C, Heger D, De~Pesters A, Telaar D, Brunner P, Schalk G, et~al.
\newblock Brain-to-text: decoding spoken phrases from phone representations in
  the brain.
\newblock Frontiers in neuroscience 2015;9:217.

\bibitem[{Moses et~al.(2016)Moses, David A and Mesgarani, Nima and Leonard,
  Matthew K and Chang, Edward F}]{moses2016neural}
Moses DA, Mesgarani N, Leonard MK, Chang EF.
\newblock Neural speech recognition: continuous phoneme decoding using
  spatiotemporal representations of human cortical activity.
\newblock Journal of neural engineering 2016;13(5):056004.

\bibitem[{Moses et~al.(2018)Moses, David A and Leonard, Matthew K and Chang,
  Edward F}]{moses2018real}
Moses DA, Leonard MK, Chang EF.
\newblock Real-time classification of auditory sentences using evoked cortical
  activity in humans.
\newblock Journal of neural engineering 2018;15(3):036005.

\bibitem[{Moses et~al.(2019)Moses, David A and Leonard, Matthew K and Makin,
  Joseph G and Chang, Edward F}]{moses2019real}
Moses DA, Leonard MK, Makin JG, Chang EF.
\newblock Real-time decoding of question-and-answer speech dialogue using human
  cortical activity.
\newblock Nature communications 2019;10(1):1--14.

\bibitem[{Madotto et~al.(2020)Madotto, Andrea and Liu, Zihan and Lin, Zhaojiang
  and Fung, Pascale}]{madotto2020language}
Madotto A, Liu Z, Lin Z, Fung P.
\newblock Language models as few-shot learner for task-oriented dialogue
  systems.
\newblock arXiv preprint arXiv:200806239 2020;.

\bibitem[{Angrick et~al.(2019)Angrick, Miguel and Herff, Christian and Mugler,
  Emily and Tate, Matthew C and Slutzky, Marc W and Krusienski, Dean J and
  Schultz, Tanja}]{angrick2019speech}
Angrick M, Herff C, Mugler E, Tate MC, Slutzky MW, Krusienski DJ, et~al.
\newblock Speech synthesis from ECoG using densely connected 3D convolutional
  neural networks.
\newblock Journal of neural engineering 2019;16(3):036019.

\bibitem[{Akbari et~al.(2019)Akbari, Hassan and Khalighinejad, Bahar and
  Herrero, Jose L and Mehta, Ashesh D and Mesgarani, Nima}]{akbari2019towards}
Akbari H, Khalighinejad B, Herrero JL, Mehta AD, Mesgarani N.
\newblock Towards reconstructing intelligible speech from the human auditory
  cortex.
\newblock Scientific reports 2019;9(1):874.

\bibitem[{Berezutskaya et~al.(2020)Berezutskaya, Julia and Freudenburg, Zachary
  V and G{\"u}{\c{c}}l{\"u}, Umut and van Gerven, Marcel AJ and Ramsey, Nick
  F}]{berezutskaya2020brain}
Berezutskaya J, Freudenburg ZV, G{\"u}{\c{c}}l{\"u} U, van Gerven MA, Ramsey
  NF.
\newblock Brain-optimized extraction of complex sound features that drive
  continuous auditory perception.
\newblock PLoS computational biology 2020;16(7):e1007992.

\bibitem[{Makin et~al.(2020)Makin, Joseph G and Moses, David A and Chang,
  Edward F}]{makin2020machine}
Makin JG, Moses DA, Chang EF.
\newblock Machine translation of cortical activity to text with an
  encoder--decoder framework.
\newblock Nature neuroscience 2020;23(4):575--582.

\bibitem[{Moses et~al.(2021)Moses, David A and Metzger, Sean L and Liu, Jessie
  R and Anumanchipalli, Gopala K and Makin, Joseph G and Sun, Pengfei F and
  Chartier, Josh and Dougherty, Maximilian E and Liu, Patricia M and Abrams,
  Gary M and others}]{moses2021neuroprosthesis}
Moses DA, Metzger SL, Liu JR, Anumanchipalli GK, Makin JG, Sun PF, et~al.
\newblock Neuroprosthesis for Decoding Speech in a Paralyzed Person with
  Anarthria.
\newblock New England Journal of Medicine 2021;385(3):217--227.

\bibitem[{Anumanchipalli et~al.(2019)Anumanchipalli, Gopala K. and Chartier,
  Josh and Chang, Edward F.}]{Anumanchipalli2019}
Anumanchipalli GK, Chartier J, Chang EF.
\newblock Speech synthesis from neural decoding of spoken sentences.
\newblock Nature 2019;568(7753):493--498.
\newblock \urlprefix\url{https://doi.org/10.1038/s41586-019-1119-1}.

\bibitem[{van~der Loo et~al.(2017)van der Loo, Lars E and Schijns, Olaf EMG and
  Hoogland, Govert and Colon, Albert J and Wagner, G Louis and Dings, Jim TA
  and Kubben, Pieter L}]{van2017methodology}
van~der Loo LE, Schijns OE, Hoogland G, Colon AJ, Wagner GL, Dings JT, et~al.
\newblock Methodology, outcome, safety and in vivo accuracy in traditional
  frame-based stereoelectroencephalography.
\newblock Acta neurochirurgica 2017;159(9):1733--1746.

\bibitem[{Herff et~al.(2020)Herff, Christian and Krusienski, Dean J and Kubben,
  Pieter}]{herff2020potential}
Herff C, Krusienski DJ, Kubben P.
\newblock The Potential of Stereotactic-EEG for Brain-Computer Interfaces:
  Current Progress and Future Directions.
\newblock Frontiers in Neuroscience 2020;14:123.

\bibitem[{Petrosyan et~al.(2022)Petrosyan, Artur and Voskoboinikov, Alexey and
  Sukhinin, Dmitrii and Makarova, Anna and Skalnaya, Anastasia and Arkhipova,
  Nastasia and Sinkin, Mikhail and Ossadtchi, Alexei}]{petrosyan2022speech}
Petrosyan A, Voskoboinikov A, Sukhinin D, Makarova A, Skalnaya A, Arkhipova N,
  et~al.
\newblock Speech decoding from a small set of spatially segregated minimally
  invasive intracranial EEG electrodes with a compact and interpretable neural
  network.
\newblock bioRxiv 2022;.

\bibitem[{Angrick et~al.(2022)Angrick, Miguel and Ottenhoff, Maarten and
  Diener, Lorenz and Ivucic, Darius and Ivucic, Gabriel and Goulis, Sophocles
  and Colon, Albert J and Wagner, Louis and Krusienski, Dean J and Kubben,
  Pieter L and others}]{angrick2022towards}
Angrick M, Ottenhoff M, Diener L, Ivucic D, Ivucic G, Goulis S, et~al.
\newblock Towards Closed-Loop Speech Synthesis from Stereotactic EEG: A Unit
  Selection Approach.
\newblock In: ICASSP 2022-2022 IEEE International Conference on Acoustics,
  Speech and Signal Processing (ICASSP) IEEE; 2022. p. 1296--1300.

\bibitem[{Angrick et~al.(2021)Angrick, Miguel and Ottenhoff, Maarten C and
  Diener, Lorenz and Ivucic, Darius and Ivucic, Gabriel and Goulis, Sophocles
  and Saal, Jeremy and Colon, Albert J and Wagner, Louis and Krusienski, Dean J
  and others}]{angrick2021real}
Angrick M, Ottenhoff MC, Diener L, Ivucic D, Ivucic G, Goulis S, et~al.
\newblock Real-time synthesis of imagined speech processes from minimally
  invasive recordings of neural activity.
\newblock Communications biology 2021;4(1):1--10.

\bibitem[{Vaswani et~al.(2017)Vaswani, Ashish and Shazeer, Noam and Parmar,
  Niki and Uszkoreit, Jakob and Jones, Llion and Gomez, Aidan N and Kaiser,
  Lukasz and Polosukhin, Illia}]{vaswani2017attention}
Vaswani A, Shazeer N, Parmar N, Uszkoreit J, Jones L, Gomez AN, et~al.
\newblock Attention is all you need.
\newblock arXiv preprint arXiv:170603762 2017;.

\bibitem[{Ardila et~al.(2019)Ardila, Rosana and Branson, Megan and Davis, Kelly
  and Henretty, Michael and Kohler, Michael and Meyer, Josh and Morais, Reuben
  and Saunders, Lindsay and Tyers, Francis M and Weber,
  Gregor}]{ardila2019common}
Ardila R, Branson M, Davis K, Henretty M, Kohler M, Meyer J, et~al.
\newblock {Common voice: A massively-multilingual speech corpus}.
\newblock arXiv preprint arXiv:191206670 2019;.

\bibitem[{Kothe(2014)Kothe, Christian}]{kothe2014lab}
Kothe C.
\newblock Lab streaming layer (LSL).
\newblock https://github com/sccn/labstreaminglayer Accessed on October
  2014;26:2015.

\bibitem[{Roussel et~al.(2020)Roussel, Phil{\'e}mon and Le Godais, Ga{\"e}l and
  Bocquelet, Florent and Palma, Marie and Hongjie, Jiang and Zhang, Shaomin and
  Giraud, Anne-Lise and M{\'e}gevand, Pierre and Miller, Kai and Gehrig,
  Johannes and others}]{roussel2020observation}
Roussel P, Le~Godais G, Bocquelet F, Palma M, Hongjie J, Zhang S, et~al.
\newblock Observation and assessment of acoustic contamination of
  electrophysiological brain signals during speech production and sound
  perception.
\newblock Journal of Neural Engineering 2020;17(5):056028.

\bibitem[{Hamilton et~al.(2017)Hamilton, Liberty S and Chang, David L and Lee,
  Morgan B and Chang, Edward F}]{hamilton2017semi}
Hamilton LS, Chang DL, Lee MB, Chang EF.
\newblock Semi-automated anatomical labeling and inter-subject warping of
  high-density intracranial recording electrodes in electrocorticography.
\newblock Frontiers in Neuroinformatics 2017;11:62.

\bibitem[{Fischl(2012)Fischl, Bruce}]{fischl2012freesurfer}
Fischl B.
\newblock FreeSurfer.
\newblock Neuroimage 2012;62(2):774--781.

\bibitem[{McFee et~al.(2015)McFee, Brian and Raffel, Colin and Liang, Dawen and
  Ellis, Daniel PW and McVicar, Matt and Battenberg, Eric and Nieto,
  Oriol}]{mcfee2015librosa}
McFee B, Raffel C, Liang D, Ellis DP, McVicar M, Battenberg E, et~al.
\newblock librosa: Audio and music signal analysis in python.
\newblock In: Proceedings of the 14th python in science conference, vol.~8;
  2015. .

\bibitem[{Shen et~al.(2018)Shen, Jonathan and Pang, Ruoming and Weiss, Ron J
  and Schuster, Mike and Jaitly, Navdeep and Yang, Zongheng and Chen, Zhifeng
  and Zhang, Yu and Wang, Yuxuan and Skerrv-Ryan, Rj and
  others}]{shen2018natural}
Shen J, Pang R, Weiss RJ, Schuster M, Jaitly N, Yang Z, et~al.
\newblock Natural tts synthesis by conditioning wavenet on mel spectrogram
  predictions.
\newblock In: 2018 IEEE International Conference on Acoustics, Speech and
  Signal Processing (ICASSP) IEEE; 2018. p. 4779--4783.

\bibitem[{Leuthardt et~al.(2012)Leuthardt, Eric and Pei, Xiao-Mei and
  Breshears, Jonathan and Gaona, Charles and Sharma, Mohit and Freudenburg,
  Zachary and Barbour, Dennis and Schalk, Gerwin}]{leuthardt2012temporal}
Leuthardt E, Pei XM, Breshears J, Gaona C, Sharma M, Freudenburg Z, et~al.
\newblock Temporal evolution of gamma activity in human cortex during an overt
  and covert word repetition task.
\newblock Frontiers in human neuroscience 2012;6:99.

\bibitem[{Crone et~al.(2001)Crone, NEea and Hao, L and Hart, J and Boatman, D
  and Lesser, RP and Irizarry, R and Gordon,
  B}]{crone2001electrocorticographic}
Crone N, Hao L, Hart J, Boatman D, Lesser R, Irizarry R, et~al.
\newblock Electrocorticographic gamma activity during word production in spoken
  and sign language.
\newblock Neurology 2001;57(11):2045--2053.

\bibitem[{Towle et~al.(2008)Towle, Vernon L and Yoon, Hyun-Ah and Castelle,
  Michael and Edgar, J Christopher and Biassou, Nadia M and Frim, David M and
  Spire, Jean-Paul and Kohrman, Michael H}]{towle2008ecog}
Towle VL, Yoon HA, Castelle M, Edgar JC, Biassou NM, Frim DM, et~al.
\newblock ECoG gamma activity during a language task: differentiating
  expressive and receptive speech areas.
\newblock Brain 2008;131(8):2013--2027.

\bibitem[{Ray et~al.(2008)Ray, Supratim and Crone, Nathan E and Niebur, Ernst
  and Franaszczuk, Piotr J and Hsiao, Steven S}]{ray2008neural}
Ray S, Crone NE, Niebur E, Franaszczuk PJ, Hsiao SS.
\newblock Neural correlates of high-gamma oscillations (60--200 Hz) in macaque
  local field potentials and their potential implications in
  electrocorticography.
\newblock Journal of Neuroscience 2008;28(45):11526--11536.

\bibitem[{Kern et~al.(2019)Kern, Markus and Bert, Sina and Glanz, Olga and
  Schulze-Bonhage, Andreas and Ball, Tonio}]{kern2019human}
Kern M, Bert S, Glanz O, Schulze-Bonhage A, Ball T.
\newblock Human motor cortex relies on sparse and action-specific activation
  during laughing, smiling and speech production.
\newblock Communications biology 2019;2(1):1--14.

\bibitem[{Prenger et~al.(2019)Prenger, Ryan and Valle, Rafael and Catanzaro,
  Bryan}]{prenger2019waveglow}
Prenger R, Valle R, Catanzaro B.
\newblock Waveglow: A flow-based generative network for speech synthesis.
\newblock In: ICASSP 2019-2019 IEEE International Conference on Acoustics,
  Speech and Signal Processing (ICASSP) IEEE; 2019. p. 3617--3621.

\bibitem[{Cho et~al.(2014)Cho, Kyunghyun and Van Merri{\"e}nboer, Bart and
  Gulcehre, Caglar and Bahdanau, Dzmitry and Bougares, Fethi and Schwenk,
  Holger and Bengio, Yoshua}]{cho2014learning}
Cho K, Van~Merri{\"e}nboer B, Gulcehre C, Bahdanau D, Bougares F, Schwenk H,
  et~al.
\newblock Learning phrase representations using RNN encoder-decoder for
  statistical machine translation.
\newblock arXiv preprint arXiv:14061078 2014;.

\bibitem[{Ioffe and Szegedy(2015)Ioffe, Sergey and Szegedy,
  Christian}]{ioffe2015batch}
Ioffe S, Szegedy C.
\newblock Batch normalization: Accelerating deep network training by reducing
  internal covariate shift.
\newblock In: International conference on machine learning PMLR; 2015. p.
  448--456.

\bibitem[{Bahdanau et~al.(2014)Bahdanau, Dzmitry and Cho, Kyunghyun and Bengio,
  Yoshua}]{bahdanau2014neural}
Bahdanau D, Cho K, Bengio Y.
\newblock Neural machine translation by jointly learning to align and
  translate.
\newblock arXiv preprint arXiv:14090473 2014;.

\bibitem[{Loshchilov and Hutter(2017)Loshchilov, Ilya and Hutter,
  Frank}]{loshchilov2017decoupled}
Loshchilov I, Hutter F.
\newblock Decoupled weight decay regularization.
\newblock arXiv preprint:171105101 2017;.

\bibitem[{Kraft and Z{\"o}lzer(2014)Kraft, Sebastian and Z{\"o}lzer,
  Udo}]{kraft2014beaqlejs}
Kraft S, Z{\"o}lzer U.
\newblock BeaqleJS: HTML5 and JavaScript based framework for the subjective
  evaluation of audio quality.
\newblock In: Linux Audio Conference, Karlsruhe, DE; 2014. .

\bibitem[{Berezutskaya et~al.(2022)Berezutskaya, Julia and Freudenburg, Zachary
  V and Vansteensel, Mariska J and Aarnoutse, Erik J and Ramsey, Nick F and van
  Gerven, Marcel AJ}]{berezutskaya2022direct}
Berezutskaya J, Freudenburg ZV, Vansteensel MJ, Aarnoutse EJ, Ramsey NF, van
  Gerven MA.
\newblock Direct Speech Reconstruction from Sensorimotor Brain Activity with
  Optimized Deep Learning Models.
\newblock bioRxiv 2022;.

\bibitem[{Brumberg et~al.(2016)Brumberg, J.S. and Krusienski, D.J. and
  Chakrabarti, S. and Gunduz, A. and Brunner, P. and Ritaccio, A.L. and Schalk,
  G.}]{brumberg2016spatio}
Brumberg JS, Krusienski DJ, Chakrabarti S, Gunduz A, Brunner P, Ritaccio AL,
  et~al.
\newblock {Spatio-Temporal Progression of Cortical Activity Related to
  Continuous Overt and Covert Speech Production in a Reading Task}.
\newblock PloS one 2016;11(11):e0166872.

\bibitem[{Hickok and Poeppel(2000)Hickok, Gregory and Poeppel,
  David}]{hickok2000towards}
Hickok G, Poeppel D.
\newblock Towards a functional neuroanatomy of speech perception.
\newblock Trends in cognitive sciences 2000;4(4):131--138.

\bibitem[{Hickok(2012)Hickok, Gregory}]{hickok2012computational}
Hickok G.
\newblock Computational neuroanatomy of speech production.
\newblock Nature reviews neuroscience 2012;13(2):135--145.

\bibitem[{van~de Ven et~al.(2020)van de Ven, Vincent and Waldorp, Lourens and
  Christoffels, Ingrid}]{van2020hippocampus}
van~de Ven V, Waldorp L, Christoffels I.
\newblock Hippocampus plays a role in speech feedback processing.
\newblock NeuroImage 2020;223:117319.

\bibitem[{Stuart et~al.(2002)Stuart, Andrew and Kalinowski, Joseph and
  Rastatter, Michael P and Lynch, Kerry}]{stuart2002effect}
Stuart A, Kalinowski J, Rastatter MP, Lynch K.
\newblock Effect of delayed auditory feedback on normal speakers at two speech
  rates.
\newblock The Journal of the Acoustical Society of America
  2002;111(5):2237--2241.

\end{thebibliography}



\end{document}